\documentclass[aps,prl,notitlepage,twocolumn,superscriptaddress,nolongbibliography,nofootinbib]{revtex4-2}
\usepackage{amsmath}
\usepackage{amssymb}
\usepackage{bbold}
\usepackage{xcolor}
\usepackage{comment}
\usepackage{upgreek}
\usepackage{graphicx}
\usepackage{xparse}
\usepackage{mathtools}
\definecolor{dodgerblue}{HTML}{1E90FF}
\usepackage[colorlinks=true,citecolor=dodgerblue,linkcolor=dodgerblue,urlcolor=dodgerblue]{hyperref}
\usepackage{physics}
\usepackage[utf8]{inputenc}

\def\d{{\rm d}}

\newcommand{\veczeta}{\vec{\zeta}\mskip2mu}
\newcommand{\vecx}{\vec{x}}
\newcommand{\vecq}{\vec{q}\mskip2mu}
\newcommand{\vecv}{\vec{v}}
\newcommand{\vecz}{\vec{z}\mskip1mu}
\newcommand{\vecp}{\vec{p}\mskip2mu}

\newcommand{\vecE}{\vec{E}}
\newcommand{\veck}{\vec{k}}

\newcommand{\hatvecx}{\hat{\vecx}}
\newcommand{\hatvecq}{\hat{\vecq}\mskip2mu}

\newcommand{\hatvecsigma}{\hat{\vec{\eta}}\mskip1mu}

\newcommand{\psiSC}{\psi}

\newcommand{\weyl}[1]{{\hat{\mathbf{#1}}}}

\newcommand{\vweyl}[1]{{\hat{\vec{\mathbf{#1}}}}}
\NewDocumentCommand{\AGP}{m o}{\weyl{A}_{#1}\IfValueT{#2}{^{#2}}}
\NewDocumentCommand{\vAGP}{o}{\vweyl{A}\IfValueT{#1}{_{#1}}}
\NewDocumentCommand{\vAGPBO}{o}{\hat{\vec{\mathbf{A}}}\IfValueT{#1}{_{#1}}^{(0)}}

\newcommand{\eBOn}{\varepsilon^{(n)}_{\rm BO}} %Born-Oppenheimer surface
\newcommand{\eBOm}{\varepsilon^{(m)}_{\rm BO}} %Born-Oppenheimer surface v2
\newcommand{\eBOnk}{\varepsilon^{(n\veck)}_{\rm BO}} %Born-Oppenheimer surface v3
\newcommand{\UBOn}{U_{\rm BO}^{(n)}} %Total BO potential energy surface
\newcommand{\psin}{\psi_{n}} %Kinematic state
\newcommand{\en}{\varepsilon^{(n)}} %Kinematic surface
\newcommand{\enk}{\varepsilon^{(n\veck)}} %Kinematic surface v2
\newcommand{\Un}{U^{(n)}} %Total kinematic energy surface

\begin{document}
\makeatletter
\newcommand{\balancecolsandclearpage}{%
    \close@column@grid
    \clearpage
    \twocolumngrid
}
\makeatother
\setcounter{secnumdepth}{2}
\title{Kinematic Quantum Geometry}
\author{Bernardo Barrera}
\affiliation{Department of Physics, Boston University, Boston, MA 02215, USA}
\author{Nishchhal Verma}
\affiliation{Department of Physics, Columbia University, New York, NY 10027, USA}
\author{Raquel Queiroz}
\affiliation{Department of Physics, Columbia University, New York, NY 10027, USA}
\affiliation{Center for Computational Quantum Physics, Flatiron Institute, New York, NY 10010, USA}
\author{Anatoli Polkovnikov}
\affiliation{Department of Physics, Boston University, Boston, MA 02215, USA}
\author{Anushya Chandran}
\affiliation{Department of Physics, Boston University, Boston, MA 02215, USA}
\begin{abstract}

Kinematic quantum geometry describes fast degrees of freedom (DOFs) that adiabatically follow states parametrized by both the coordinates and the velocities of slow DOFs. It extends Born-Oppenheimer geometry, which depends on coordinates alone, introduces mixed Berry curvatures, and modifies the effective theory governing the slow DOFs. We test its predictions in two setups: spinful particles moving in magnetic textures, and Dirac electrons in graphene coupled to an LC circuit. In the first, kinematic quantum geometry produces a velocity-dependent effective mass tensor and generates a  Lorentz-like force nonlinear in velocity that can arise even in coplanar textures. In the second, it renormalizes the circuit capacitance and generates a Hall current nonlinear in the electric field. For small velocities, it reproduces recent results for electronic dynamics in external electromagnetic fields and in pseudospin textures. However, the framework can remain valid at large velocities beyond the perturbative regime, provided the kinematic states are sufficiently slowly varying.
\end{abstract}

\maketitle

\paragraph*{Introduction.---} 
The Born-Oppenheimer (BO) approximation is a foundational framework for systems with a clear timescale separation between slow and fast degrees of freedom (DOFs): the fast DOFs adiabatically follow an instantaneous eigenstate $\ket{\smash{\psin^{\rm BO}(\vecx)}}$ determined by a set of slow coordinates $\vecx$~\cite{BornOppenheimer1927}. It underlies molecular dynamics~\cite{atkins2011molecular} and condensed matter physics, from first-principles electron-phonon calculations~\cite{baroni2001phonons,giustino2017electron} to the semiclassical dynamics of Bloch electrons in external fields~\cite{ashcroft1976solid,sundaram1999wave}.

The quantum geometry of the BO manifold of states affects the dynamics of the slow DOFs. It is encoded in the quantum geometric tensor, whose real and imaginary parts give the quantum metric and the Berry curvature $\mathcal{F}^{(n)}_{x_ix_j}$, respectively~\cite{Provost1980RiemannianStates}. Early work by Mead and Truhlar~\cite{mead1979determination}, and Berry~\cite{berry1989quantum,berry1993classical}, showed that the Berry curvature acts as an effective magnetic field on the slow DOFs, while the metric corrects their effective potential $\UBOn(\vecx)$. Both enter the equation of motion for the slow DOFs, which have mass $M$,
\begin{equation}
M\ddot{x}_i = -\partial_{x_i}\UBOn +\dot{x}_j\mathcal{F}^{(n)}_{x_ix_j}.
\label{eq:adiabatic_eom}
\end{equation}
The first term on the right-hand side of Eq.~\eqref{eq:adiabatic_eom} is a gradient force, whereas the second is a Lorentz force arising from the Berry curvature.

Berry curvature effects have been studied extensively in electronic systems, where they explain several Hall responses. In the topological Hall effect, the slow coordinate is the electronic position, and the Berry curvature induced by a noncoplanar magnetic texture deflects electron trajectories in real space~\cite{bruno2004topological,neubauer2009topological,yu2020giant}. In the anomalous Hall effect, the slow coordinate is the crystal momentum, and the Berry curvature of Bloch wavefunctions generates a current transverse to the electric field~\cite{nagaosa2010anomalous,xiao2010berry}.

\begin{figure}
    \centering
    \includegraphics[width = 0.9\columnwidth]{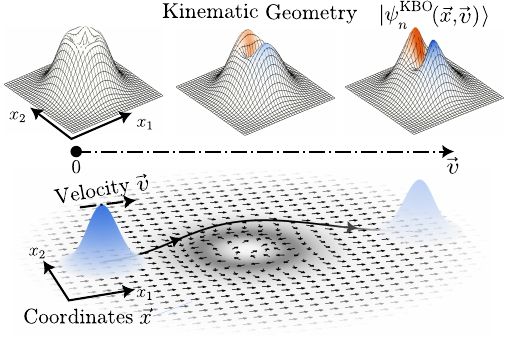}
    \caption{\textit{Kinematic quantum geometry.} 
    Lower panel: A wavepacket, characterized by a set of general coordinates $\vecx$ and velocity $\vecv$, moves through a region where the fast states have non-trivial kinematic quantum geometry. The arrows represent the local BO state $\ket{\psin^{\rm BO}(\vecx)}$ as a spinor, which rotates significantly in the shaded region. 
    Upper panel: Successive $(x_1,x_2)$ slices of the kinematic geometry at increasing velocity $\vecv$; the BO geometry is recovered at $\vecv=0$. 
    The surface height shows the trace of the kinematic metric $\operatorname{Tr}\boldsymbol{h}^{(n)}_{xx}$, while the color shows the kinematic Berry curvature $\mathcal{K}_{x_1x_2}$.
    }
    \label{fig:main1}
\end{figure}

In this work, we formulate quantum geometry beyond the BO limit. When the BO states change significantly as the slow DOFs move, the fast DOFs do not adiabatically follow these states. Instead, they adiabatically follow a family of \emph{kinematic} Born-Oppenheimer states $\ket{\psin^{\rm KBO}(\vecx,\vecv)}$, which depend on both the coordinates $\vecx$ and their instantaneous velocities $\vecv$. 
The kinematic states are eigenstates of the Hamiltonian in a frame comoving with the BO basis, build in pseudoforces that the motion of the slow DOFs exerts on the fast DOFs, and have been brought into focus by Refs.~\cite{Shenvi2009,barrera2026MBOA,bian2025phase,bian2026phase}. They coincide with Berry's first superadiabatic states for given $\vecx(t)$~\cite{berry1987quantum}. We refer to their quantum geometry on the expanded space of both coordinates and velocities as \emph{kinematic quantum geometry} (KQG), see Fig.~\ref{fig:main1}. 

We show that KQG describes beyond-BO dynamics in systems of coupled slow and fast DOFs. In particular, the slow DOFs obey the \emph{kinematic} equation of motion,
\begin{equation}
    \left(M\delta_{ij}-{\mathcal{K}}_{x_iv_j}^{(n)}\right)\, \ddot{x}_j=
    -\partial_{x_i}\Un_{\rm KBO} +\dot{x}_j{\mathcal{K}}_{x_ix_j}^{(n)}.
    \label{eq:kinematic_eom}
\end{equation}
Here, the spatial Berry curvature $\mathcal{K}^{(n)}_{x_ix_j}$ and the effective potential $\Un_{\rm KBO}$, play the roles of $\mathcal{F}^{(n)}_{x_ix_j}$ and $\UBOn$ in Eq.~\eqref{eq:adiabatic_eom}, but depend on both $\vecx$ and $\vecv=\dot{\vecx}$ (Fig.~\ref{fig:main1}). The new ingredient is the mixed $(x$-$v)$ Berry curvature $\mathcal{K}^{(n)}_{x_iv_j}$, which has no BO counterpart, and gives a renormalized velocity-dependent mass tensor for the slow DOFs.

By comparing to numerical simulations, we demonstrate that KQG determines dynamics in two model systems: (i) a spinful particle moving through a coplanar magnetic texture and (ii) Dirac electrons in graphene coupled to an LC circuit. In both cases, BO geometry has no local Berry curvature, while KQG is non-trivial. In the first, KQG produces a velocity-dependent mass tensor for the particle, and generates a Lorentz-like force that deflects its real space trajectory. In the second, KQG develops a Berry curvature dipole, generating a Hall response that is non-linear in the electric field. It also renormalizes the capacitance of the circuit, producing a shift in its oscillation frequency. 
Beyond-BO effects, captured by Eq.~\eqref{eq:kinematic_eom}, are large when (i) the texture varies rapidly in space, and (ii) the circuit’s electric field is strong, respectively.

Beyond-BO corrections in both systems have been studied before, but separately and perturbatively in $\vec{v}$. For magnetic textures, recent work has interpreted them as emergent gravity~\cite{onishi2026emergent,yoshida2025emergent}. For Bloch electrons in electromagnetic fields, the corrections are organized perturbatively in the fields. This program dates back to Blount~\cite{blount1962bloch} and has evolved into a larger modern literature~\cite{gao2014field,gao2019semiclassical,kaplan2024unification} on non-linear response~\cite{lai2021third,liu2022berry}.
Eq.~\eqref{eq:kinematic_eom} shows that phenomena in both systems are manifestations of a single object, KQG, and describes them non-perturbatively in $\vec{v}$. 

\paragraph{Kinematic Quantum Geometry.---}
Consider a Hamiltonian $\weyl{H}_{\rm int}(\vecx)$ acting on a set of fast degrees of freedom, and parametrized by a set of slow coordinates $\vecx$~\footnote{Throughout, bold hatted symbols denote operators acting on the fast DOFs with the slow ones treated as fixed parameters (numbers). Formally, this operation corresponds to a partial Wigner-Weyl transform; see End Matter.}. 

BO asserts that the fast DOFs adiabatically follow the instantaneous eigenstates $\ket{\psin^{\rm BO}(\vecx)}$ of the Hamiltonian $\weyl{H}_{\rm int}(\vecx)$. 
The BO adiabatic gauge potential (AGP),
\begin{equation}
    i\hbar\partial_{x_j}\ket{\psin^{\rm BO}(\vecx)} = \weyl{A}_{x_j}^{\rm BO}(\vecx)\ket{\psin^{\rm BO}(\vecx)},
    \label{eq:BO_AGP}
\end{equation}
is the generator of adiabatic transformations of the BO states under variations of the slow coordinates. 

When the rate of change of the slow coordinates is significant, 
we assert that the fast DOFs adiabatically follow a new family of \emph{kinematic states} $\ket{\psin^{\rm KBO}(\vecx,\vecv)}$, which are eigenstates of the Hamiltonian,
\begin{equation}
    \weyl{H}_{\rm int}^{\rm K}(\vecx,\vecv) = \weyl{H}_{\rm int}(\vecx)-v_j\weyl{A}^{\rm BO}_{x_j}(\vecx).
    \label{eq:H_K}
\end{equation}
Throughout, summation over repeated indices is implied. The Hamiltonian in Eq.~\eqref{eq:H_K} describes the fast DOFs in a frame co-moving with the BO basis~\cite{kolodrubetz2017geometry}, where the additional term $-v_j\weyl{A}^{\rm BO}_{x_j}$ is responsible for pseudoforces arising from motion.
The kinematic states are thus analogues of mechanical equilibrium in a moving frame~\cite{barrera2026MBOA}, and can differ substantially from laboratory-frame equilibrium; for example, the kinematic state of a gas in a rotating box carries angular momentum. 

The expectation values of the Hamiltonian,
\begin{equation}
    \en_{\rm KBO}(\vecx,\vecv) \equiv \langle \weyl{H}_{\rm int}(\vecx)\rangle_n,
    \label{eq:KBO_surface}
\end{equation}
define the \emph{kinematic energy surfaces}. Throughout, $\langle\weyl{O}\rangle_n \equiv \bra{\smash{\psi_n^{\rm KBO}}}\weyl{O}\ket{\smash{\psi_n^{\rm KBO}}}$  denotes an expectation value in a kinematic state. For vanishing velocity, Eq.~\eqref{eq:KBO_surface} reduces to the usual BO surface: $\en_{\rm BO}(\vecx)=\en_{\rm KBO}(\vecx,\vecv=0)$. Note that $\en_{\rm KBO}$ is the expectation value of the lab frame Hamiltonian $\weyl{H}_{\rm int}$, and \emph{not} $\weyl{H}^{\rm K}_{\rm int}$. 

The AGPs associated to the kinematic states,
\begin{equation}
i\hbar\partial_{z_\alpha}\ket{\smash{\psin^{\rm KBO}(\vecz)}}
=
\weyl{A}_{\alpha}^{\rm KBO}(\vecz)\ket{\smash{\psin^{\rm KBO}(\vecz)}},
\label{eq:kinematic_AGPs}
\end{equation}
define the \emph{kinematic quantum geometry} (KQG). Above, we have defined the extended coordinates $z_\alpha=(x_i,v_i)$. Greek (Latin) indices label the extended (original) coordinates. The real and imaginary parts of the connected AGP covariance, $Q_{\alpha\beta}^{(n)}
=\langle\weyl{A}_{\alpha}^{\rm KBO}\weyl{A}_{\beta}^{\rm KBO}\rangle_{n,c}$, respectively give the kinematic quantum metric and Berry curvature,
\begin{equation}
    h_{\alpha\beta}^{(n)}=\operatorname{Re}Q_{\alpha\beta}^{(n)},
    \qquad
    \mathcal{K}_{\alpha\beta}^{(n)}
    =-\frac{2}{\hbar}\operatorname{Im}Q_{\alpha\beta}^{(n)},
    \label{eq:kinematic_geometry}
\end{equation}
where $\langle\weyl{A}\weyl{B}\rangle_{n,c}
=\langle\weyl{A}\weyl{B}\rangle_n
-\langle\weyl{A}\rangle_n\langle\weyl{B}\rangle_n$. Bold font denotes the kinematic Berry curvatures as matrices, with elements $[\boldsymbol{\mathcal{K}}^{(n)}_{xx}]_{ij}=\mathcal{K}^{(n)}_{x_i x_j}$ and $[\boldsymbol{\mathcal{K}}^{(n)}_{xv}]_{ij}=\mathcal{K}^{(n)}_{x_i v_j}$. Again, the spatial kinematic Berry curvature reduces to the BO curvature at zero velocity: $\mathcal{F}^{(n)}_{x_ix_j}(\vecx)=\mathcal{K}^{(n)}_{x_ix_j}(\vecx,\vecv=0).$ 

\begin{figure}
\centering
\includegraphics[width = \columnwidth]{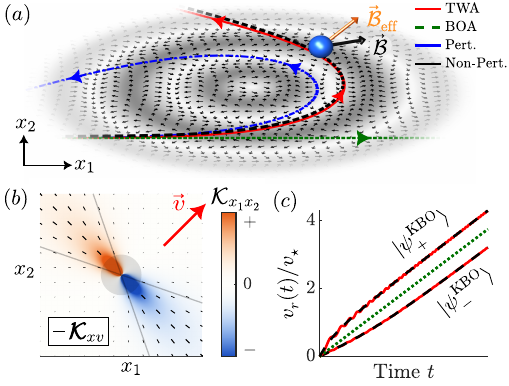}

\caption{\textit{Beyond-BO dynamics in a pseudospin texture.}
(a) A spinful particle moves through a coplanar texture $\vec{\mathcal{B}}(\vecx)$ (black arrows). The curves show trajectories obtained from TWA, BO, leading (perturbative) correction to BO, and the (nonperturbative) kinematic equation Eq.~\eqref{eq:kinematic_eom}.
(b) Kinematic Berry curvatures. Black rods indicate the orientation and eigenvalue magnitude of the non-zero eigenvector of $\boldsymbol{\mathcal{K}}_{xv}$. Color denotes $\mathcal{K}_{x_1x_2}$, which develops a dipole at finite $\vecv$.
(c) Evolution of the radial velocity $v_r(t)$ under a constant external radial force, illustrating mass renormalization. The dotted green line shows the BO prediction $v_r(t) = Ft/M$. 
}
\label{fig:main2}
\end{figure}

A few comments about Eq.~\eqref{eq:H_K} are in order. First, the BO AGP is defined only up to a choice of gauge. For all examples considered here, we use the Kato AGP~\cite{kato1950adiabatic,schindler2025geometric} (parallel-transport law in Ref.~\cite{berry1987quantum}), which has vanishing diagonal elements in the BO basis. Second, the adiabatic following of kinematic states is controlled by a superadiabatic criterion (see Sec.~\ref{sec:superadiabatic_criterion} of End Matter), which is non-perturbative in $\vecv$~\cite{berry1987quantum}. If the BO AGP is slowly-varying in space, then adiabatic following of kinematic states can hold up to very large velocities, provided the acceleration is small (see Fig.~\ref{fig:main2}(c)).

\paragraph*{Coupled Dynamics of Slow and Fast Degrees of Freedom.---}
\label{sec:coupled_dynamics}

We now promote the slow coordinates to dynamical variables with mass $M$ and consider the time evolution of the coupled system. The fast DOFs obey the Schr\"odinger equation, 
\begin{equation}
    i\hbar\partial_t\ket{\Psi(t)} = \weyl{H}_{\rm int}(\vecx(t))\ket{\Psi(t)},
    \label{eq:TWA_fast}
\end{equation}
and exert a backreaction on the slow ones through,
\begin{equation}
    M\ddot{x}_i = -\partial_iV(\vecx) - \bra{\Psi(t)}\partial_{x_i}\weyl{H}_{\rm int}(\vecx)\ket{\Psi(t)}.
    \label{eq:TWA_slow}
\end{equation}
The total force in Eq.~\eqref{eq:TWA_slow} is the sum of that from an external potential $V$ acting only on the slow DOFs, and the interaction force derived from $\weyl{H}_{\rm int}$. Eqs.~\eqref{eq:TWA_fast} and~\eqref{eq:TWA_slow} describe a mixed quantum-classical evolution, which can be understood within a Truncated Wigner Approximation (TWA)~\cite{polkovnikov2010phase}. These equations serve as a reference against which to compare KQG predictions. More generally, the slow DOFs need not be treated classically and can instead be treated quantum mechanically, as in Ref.~\cite{barrera2026MBOA}.

We assume that, as the slow DOFs evolve through some path $\vecx(\tau)$, the fast ones adiabatically follow the instantaneous kinematic state $\ket{\psin^{\rm KBO}(\vecx(\tau),\vecv(\tau))}$, with $\vecv(\tau) = \dot{\vecx}(\tau)$. 
The kinematic states self-consistently determine the evolution of $\vecx(\tau)$ through an effective action, derived from the time-dependent variational principle (see End Matter), whose extremization yields the kinematic EOM in Eq.~\eqref{eq:kinematic_eom}, with,
\begin{equation}
    \Un_{\rm KBO}(\vecx,\vecv) \equiv \en_{\rm KBO}(\vecx,\vecv) + V(\vecx).
    \label{eq:total_surface}
\end{equation}

It is instructive to understand Eq.~\eqref{eq:kinematic_eom} in the limit of small velocities, when the kinematic states are expanded perturbatively about the BO states. Expanding Eq.~\eqref{eq:H_K} to second order in $\vecv$ and keeping terms through second order in time derivatives leads to dynamics with an effective 
mass tensor $({M}_{\rm eff}^{(n)})_{ij} = M\delta_{ij}+\kappa^{(n)}_{ij}$, where,
\begin{equation}
\kappa^{(n)}_{ij} \equiv 2\textrm{Re}
\sum\limits_{m\neq n}
\frac{
\bra{\psin^{\rm BO}}
\weyl{A}_{x_i}^{\rm BO}
\ket{\psi_m^{\rm BO}}
\bra{\psi_m^{\rm BO}}
\weyl{A}_{x_j}^{\rm BO}
\ket{\psin^{\rm BO}}
}
{\eBOm-\eBOn}.
\label{eq:mass_tensor}
\end{equation}
 Eq.~\eqref{eq:kinematic_eom} then takes the form of a forced geodesic equation, 
\begin{equation}
    ({M}_{\rm eff}^{(n)})_{ij}\,\ddot{x}_j = -\partial_{x_i}\Un_{\rm BO} + \dot{x}_j\mathcal{F}_{x_i x_j}^{(n)} - \Gamma^{(n)}_{ijk}\dot{x}_j\dot{x}_k,
    \label{eq:perturbative_eom}
\end{equation}
with $\Gamma^{(n)}_{ijk} = \frac{1}{2}(
	\partial_{x_j}\kappa^{(n)}_{ki}
	+\partial_{x_k}\kappa^{(n)}_{ji}
	-\partial_{x_i}\kappa^{(n)}_{jk})$
the Christoffel symbols associated with the mass renormalization $\kappa^{(n)}_{ij}$~\cite{d2014emergent}. This analogy to gravity has been recently derived for spinful particles moving in magnetic textures~\cite{onishi2026emergent,yoshida2025emergent} and for Bloch wavepacket motion~\cite{ren2025nonadiabaticWP,ren2026nonadiabatic,maranzana2026semiclassical}, where $\kappa_{ij}$ is known (up to normalization conventions) as the band-renormalized metric or Berry connection polarizability. 

A related framework, termed the moving Born--Oppenheimer approximation (MBOA), was recently developed by some of the authors~\cite{barrera2026MBOA}. In addition to kinematic effects, the MBOA accounts for correlated fluctuations of the velocity $\vecv$. These fluctuations give rise to two effects: (i) the metric correction to the BO energy surface and (ii) effective interactions between fast DOFs. These effects are suppressed at large mass $M$, and we neglect them in this work. In the End Matter, we further justify this approximation and derive the KBO states as the mean-field limit of the MBOA.

\textit{Dynamics in Pseudospin Textures.---}
Consider a particle of spin $S$ moving through the frozen ``radial spiral'' texture of Ref.~\cite{onishi2026emergent},
shown in Fig.~\ref{fig:main2}(a),
\begin{equation}
    \weyl{H}_{\rm int}(\vecx)=\vec{\mathcal{B}}(\vecx)\cdot\hat{\vec{\mathbf{S}}}/S,
    \quad
    \vec{\mathcal{B}}(\vecx)=J(\cos kr,\sin kr,0),
    \label{eq:H_int_spin}
\end{equation}
where $\vecx=(x_1,x_2)$ is the particle's physical position in two dimensions, with $r=|\vecx|$, and $V(\vecx)=0$. The spin is the fast DOF. We focus on the two maximally polarized branches, labeled $(+)$ and $(-)$, and take $S\gg 1$ at fixed $\hbar S$. The large-spin limit suppresses velocity fluctuations and guarantees that the TWA is close to exact.  

The BO states are parallel or antiparallel to $\vec{\mathcal{B}}(\vecx)$. Their energy surfaces are constant, $\varepsilon_{\rm BO}^{(\pm)}=\pm J$, and their spatial
Berry curvature vanishes, $\mathcal{F}_{x_1x_2}^{\pm}=0$. Thus, by application of Eq.~\eqref{eq:adiabatic_eom}, the BO predicts the particle moves freely $M\ddot{x}_i=0$, with no forces exerted by the texture.

The kinematic states are maximally polarized along an effective velocity-dependent field $\vec{\mathcal{B}}_{\rm eff}(\vecx,\vecv)$. Using the BO Kato AGP $\weyl{A}_{x_j}^{\rm BO}=\hbar k e_{r,j}\weyl{S}_z$, where $\vec e_r =\vecx/r$ is the unit radial vector, Eq.~\eqref{eq:H_K} gives,
\begin{align}
    \vec{\mathcal{B}}_{\rm eff}(\vecx,\vecv)
    &=\vec{\mathcal{B}}(\vecx)-\hbar Skv_r\hat z.
    \label{eq:radial_spiral_effective_field}
\end{align}
Motion therefore makes the spin cant out of the plane of the texture, as illustrated in Fig.~\ref{fig:main2}(a).
The canting angle $\phi = \tan^{-1}(v_r/v_\star)$, where $v_\star\equiv J/(\hbar Sk)$, is controlled by the radial velocity $v_r = \vecv\cdot\vec{e}_r$.

The canting of the spin generates non-zero kinematic Berry curvatures (see Fig.~\ref{fig:main2}(b) and End Matter),
\begin{align}
    \mathcal{K}_{x_i v_j}^{\pm}
    &=\pm\mu(v_r)e_{r,i}e_{r,j},
    \qquad
    \mathcal{K}_{v_i v_j}^{\pm}=0,
    \label{eq:radial_spiral_mixed_curvature}\\
    \mathcal{K}_{x_1x_2}^{\pm}
    &=\pm\mu(v_r)\frac{v_\theta}{r},
    \qquad
    \mu(v_r)=\frac{\hbar^2S^2k^2J^2}{|\vec{\mathcal{B}}_{\rm eff}|^3}.
    \label{eq:radial_spiral_spatial_curvature}
\end{align}
Note that the radial $v_r$ and azimuthal $v_\theta$ components of $\vec{v}$ generally vary with position, even at fixed Cartesian velocity. The kinematic energy surfaces are also velocity dependent:
\mbox{$\varepsilon_{\rm KBO}^{(\pm)}(\vecx,\vecv) =\pm  J^2/{|\vec{\mathcal{B}}_{\rm eff}(\vec x,\vec v)|}$}.

The mixed curvature $\mathcal{K}_{x_iv_j}$ renormalizes the particle's response to radial forces, while the mass controlling azimuthal acceleration is unchanged.
For purely radial motion under a constant force, the kinematic EOM gives $(M\mp\mu(\dot{r}))\ddot r=F$.
Figure~\ref{fig:main2}(c) shows the evolution of the radial velocity $v_r(t)$ for a particle initially at rest, with
$\mu(0)<M$. The parallel (excited) branch accelerates more rapidly
than the BO prediction, and the anti-parallel branch more
slowly. As $|v_r|/v_\star$ increases, $\mu(v_r)$ decreases
and the bare inertial mass $M$ is recovered in the limit $|v_r|\gg v_\star$.

The spatial curvature $\mathcal{K}_{x_1x_2}$ produces a transverse force which,
together with the kinematic energy surface, controls the deflection of incoming trajectories
(Fig.~\ref{fig:main2}(a)).
At small velocities, Eq.~\eqref{eq:kinematic_eom} reduces to the forced geodesic equation in Eq.~\eqref{eq:perturbative_eom}. For the parameters shown, however, the perturbative description substantially overestimates the deflection,
whereas the full kinematic EOM closely follows the TWA benchmark. 

\paragraph*{Bloch Electrons Coupled to an LC Circuit.---}

As a second example, consider a periodic two-dimensional crystal with equal side lengths $\ell_1=\ell_2=\ell$, threaded by two fluxes $\Phi_1$ and $\Phi_2$. The fluxes are dynamical degrees of freedom associated with an LC circuit and interact with the crystal via,
\begin{equation}
    \weyl{H}_{\rm int}(\vec{\Phi})
    =
    \sum_{\veck}\sum_{a,b} H_{ab}\left(\veck-\frac{e}{\ell \hbar}\vec{\Phi}\right)\,
    \hat{c}^{\dagger}_{a\veck}\,
    \hat{c}^{\vphantom{\dagger}}_{b\veck},
    \label{eq:H_int_bands}
\end{equation}
where $\hat{c}_{a\veck}$ annihilates an electron in orbital $a$ with quasi-momentum $\veck$. The mixed quantum-classical TWA evolution, given by Eqs.~\eqref{eq:TWA_fast} and~\eqref{eq:TWA_slow}, with $x_i=\Phi_i$, $M=C$ and $V=\vec{\Phi}^2/2L$, governs the flux dynamics,
\begin{equation}
    C\ddot{\Phi}_i + \frac{\Phi_i}{L} = {I}^{\rm el}_i(t), \quad I^{\rm el}_i = -\langle \Psi(t)|\partial_{\Phi_i}\weyl{H}_{\rm int}|\Psi(t)\rangle. 
    \label{eq:TWA_circuit}
\end{equation}

Concretely, consider doped massless graphene. Initially, the circuit fluxes are off, $\vec{\Phi}=0$, and the electrons are in equilibrium in a zero-temperature Fermi-Dirac distribution $f_{n\veck}$ with Fermi energy $\varepsilon_F>0$. Close to a Dirac point the single-particle Hamiltonian is given by $H(\veck) = \hbar v_D(k_1\sigma_x-sk_2\sigma_y) \equiv h(\veck)\cdot\vec{\sigma}$, with valley chirality $s=\pm 1$ and Dirac velocity $v_D$. 

Diagonalizing the single-particle Hamiltonian $H(\vec{\upkappa})$ at each $\vec{\upkappa}(\vec{\Phi};\vec{k}) = \veck-e\vec{\Phi}/(\ell\hbar)$ gives the BO electronic bands. The BO energy surfaces $\varepsilon^{(n\veck)}_{\rm BO}(\vec{\Phi})$ are band dispersions, with corresponding flux Berry curvature $\mathcal{F}^{(n\veck)}_{\Phi_1\Phi_2}$. BO asserts that the electronic current is related to the instantaneous circuit variables as $\vec{I}^{\rm el}_{\rm BO} = \sum_{n\veck} f_{n\veck} \vec{\mathcal{J}}^{(n\veck)}_{\rm BO}$, where,
\begin{align}
    \vec{\mathcal{J}}^{(n\veck)}_{\rm BO} &= -\vec{\nabla}_{\Phi}\enk_{\rm BO} + \boldsymbol{\mathcal{F}}^{(n\veck)}_{\Phi\Phi}\cdot\dot{\vec{\Phi}}.
    \label{eq:BO_current}
\end{align}
The first term in Eq.~\eqref{eq:BO_current} is the inductive response of the crystal. The second term is the Hall response, which produces a current perpendicular to the instantaneous electric field $\vecE = -\dot{\vec{\Phi}}/\ell$. For massless graphene, the Hall contribution vanishes as inversion and time-reversal symmetry imply that $\boldsymbol{\mathcal{F}}_{\Phi\Phi}^{(n\veck)}=0$. A Taylor expansion for small flux thus gives $\vec{I}_{\rm BO}^{\rm el} = -\boldsymbol{L}^{-1}_{\rm el}\cdot\vec{\Phi} +\mathcal{O}(\vec{\Phi}^3)$, where $\boldsymbol{L}_{\rm el}^{-1}$ is an electronic renormalization of the circuit inductance proportional to the Drude weight or charge stiffness~\cite{resta2018drude}.

\begin{figure}
    \centering
    \includegraphics[]{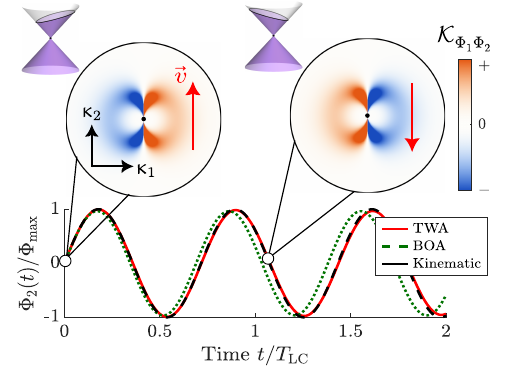}
    \caption{\textit{Coupled dynamics of LC circuit and Dirac electrons in graphene.} The lower panel shows the time evolution of $\Phi_2(t)$, demonstrating a renormalization of the LC oscillation frequency. The insets show the upper-band kinematic Berry curvature $\mathcal{K}_{\Phi_1\Phi_2}^{(+\veck)}$ near the Dirac points at different instants of time. The kinematic Berry curvature develops a dipole in the direction orthogonal to the instantaneous circuit voltage $\vecv$, shown as red arrows. For a circularly polarized initial condition, the tilt of the Fermi distribution couples to this dipole, producing a transverse Hall current that is non-linear in the electric field.
    }
    \label{fig:main3}
\end{figure}

The kinematic bands are found by diagonalizing the single-particle moving Hamiltonian (see End Matter),
\begin{equation}
    H^{\rm K}(\vec{\upkappa},\vecv) = H(\vec{\upkappa}) + \frac{e{v}_j}{\ell \hbar}A^{\rm BO}_{{\upkappa}_j}(\vec{\upkappa}),
\end{equation}
at each $\vec{\upkappa}(\vec{\Phi};\vec{k}) = \veck-e\vec{\Phi}/(\ell\hbar)$ and $\vecv$. Above, $A^{\rm BO}_{\upkappa_j}(\vec{\upkappa})$ is the single-particle AGP associated to $H(\vec{\upkappa})$, which for graphene reads $A^{\rm BO}_{\upkappa_j} = -s\hbar\sigma_z e_{\theta,j}/(2|\vec{\upkappa}|)$, where $\vec{e}_\theta = \hat{z}\times\vec{e}_{\upkappa}$ and $\vec{e}_{\upkappa}=\vec{\upkappa}/|\vec{\upkappa}|$. 
Note that the flux velocity is the voltage across the circuit, proportional to the electric field $\vec{E}=-\vecv/\ell$ in the crystal. By analogy to Eq.~\eqref{eq:radial_spiral_effective_field}, we rewrite the moving Hamiltonian as $H^{\rm K}\left(\vec{\upkappa},\vec{v}\right) = \vec{h}_{\rm eff}(\vec{\upkappa},\vec{v})\cdot\vec{\sigma}$, where,
\begin{equation}
    \vec{h}_{\rm eff}(\vec{\upkappa},\vecv)=\vec{h}(\vec{\upkappa}) - \frac{se(\vec{v}\cdot\vec{e}_\theta)}{2\ell|\vec{\upkappa}|}\hat{z}.
\end{equation}
The kinematic dispersions are 
$\varepsilon_{\rm KBO}^{(\pm\veck)}=\pm{(\hbar v_D|\vec{\upkappa}|)^2}/{|\vec{h}_{\rm eff}|}$.

The KQG is characterized by the Berry curvatures,
\begin{align}
    \mathcal{K}^{(\pm\veck)}_{\Phi_i v_j}
    &=\pm\gamma\,e_{\theta,i}e_{\theta,j},
    \qquad
    \mathcal{K}^{(\pm\veck)}_{v_i v_j}=0,\\
    \mathcal{K}^{(\pm\veck)}_{\Phi_1\Phi_2}
    &=\mp\frac{2e(\vecv\cdot\vec e_\theta)}
    {\ell\hbar|\vec{\upkappa}|}\gamma,
    \qquad
    \gamma=\frac{e^2(\hbar v_D)^2}
    {4\ell^2|\vec{h}_{\rm eff}|^3}.
\end{align}
These determine the instantaneous electronic current as $\vec I_{\rm KBO}=\sum_{n\vec k}f_{n\vec k}\vec{\mathcal J}_{\rm KBO}^{(n\vec k)}$, where,
\begin{align}
    \vec{\mathcal{J}}^{(n\veck)}_{\rm KBO} &= -\vec{\nabla}_{\Phi}\enk_{\rm KBO} + \boldsymbol{\mathcal{K}}^{(n\veck)}_{\Phi\Phi}\cdot\dot{\vec{\Phi}} +\boldsymbol{\mathcal{K}}^{(n\veck)}_{\Phi v}\cdot\ddot{\vec{\Phi}}.
    \label{eq:KBO_current}
\end{align}
Here, $\vec\nabla_\Phi$ is taken at fixed $\vec k$ and $\vec v$, with all quantities subsequently evaluated at $\vec v=\dot{\vec\Phi}$. 

The mixed curvature $\mathcal{K}_{\Phi_iv_j}$ renormalizes the capacitance of the circuit as $(C_{\rm eff})_{ij} \equiv C\delta_{ij}-\sum_{n\vec k}f_{n\vec k}\mathcal K_{\Phi_i v_j}^{(n\vec k)}$. Fig.~\ref{fig:main3} shows dynamics for an experiment where the circuit is driven out of equilibrium with initial condition $\vec{\Phi}(0)=(\Phi_0,0)$ and $\dot{\vec{\Phi}}(0)=(0,\omega_{\rm LC}\Phi_0)$ which, in the absence of electronic backreaction, would result in a circularly polarized orbit in the $\Phi_1$-$\Phi_2$ plane with frequency $\omega_{\rm LC}=1/\sqrt{LC}$. The KQG of the crystal produces an additional frequency shift relative to the BO prediction.

The flux curvature $\mathcal{K}_{\Phi_i\Phi_j}$ generates a contribution to the current in the direction orthogonal to the electric field. The insets of Fig.~\ref{fig:main3} illustrate the dipole moment in the curvature perpendicular to $\vecv = -\vec{E} \ell$. The displacement of the Fermi surface induced by the circuit flux couples to this dipole, analogously to the Berry-curvature-dipole mechanism~\cite{sodemann2015quantum}, producing a non-linear Hall response. Here, however, the kinematic Berry-curvature dipole is itself induced by the electric field, and as such the resulting Hall current appears at third order, rather than second order, in the electric field. Related third-order nonlinear Hall responses have been studied theoretically~\cite{liu2022berry} and probed experimentally through higher-harmonic AC transport measurements~\cite{lai2021third}.

\paragraph{Discussion.---}
Kinematic quantum geometry is the quantum geometry of states that depend on slow coordinates and their velocities. It organizes diverse beyond-BO phenomena, and reshapes slow dynamics (Eq.~\eqref{eq:kinematic_eom}) even where BO geometry is trivial.

The formalism remains controlled even at large velocities, far beyond the reach of perturbation theory around BO, provided the BO adiabatic gauge potential (AGP) varies slowly in space and the acceleration is small. The kinematic states then follow the eigenstates of $v_i\weyl{A}_{x_i}^{\rm BO}$ rather than those of the Hamiltonian. In Fig.~\ref{fig:main2}(b), for instance, adiabatic following of kinematic states holds at arbitrarily large radial velocity, where the spin is fully canted out of the texture plane, as long as the force on the particle, and thus the acceleration, is small.

Several directions for future work are interesting. One is to incorporate transitions between kinematic states to capture dissipation. Another is to extend the framework to interacting many-body systems, where the AGP may not be known exactly but can instead be obtained variationally~\cite{kolodrubetz2017geometry}. This will be important to understand beyond-BO physics in condensed matter settings, like coupled systems of electrons and phonons with limited time-scale separation in their dynamics, and to understand magnetic and chiral effects in molecular systems~\cite{tao2024can}, where closely related phase-space constructions are currently being developed~\cite{bian2026phase}.

\paragraph*{Acknowledgments.---}
B.B. is grateful for useful discussions with Nadya Fishchenko and Tan Nguyen. B.B. was supported by
AFOSR Grant FA9550-21-1-0342, NSF Grant
DMR-2412542, and ONR N00014-26-1-2568. N.V. and R.Q. were supported by the Air Force Office of Scientific Research under award number FA9550-26-1-B299. The Flatiron Institute is a division of the Simons Foundation.

\bibliography{refs}
\par\onecolumngrid
\clearpage
\twocolumngrid
\appendix
\begin{center}
    \Large\bfseries End Matter
\end{center}
\twocolumngrid
We derive the kinematic equation of motion Eq.~\eqref{eq:kinematic_eom}, which is the main result of this work. Sec.~\ref{sec:self_consistent_framework} introduces a general self-consistent framework for the coupled dynamics of slow and fast DOFs.  Sec.~\ref{sec:kinematic_ansatz} uses the kinematic states as an ansatz for solving the self-consistent equations, while Sec.~\ref{sec:superadiabatic_criterion} discusses the regimes of validity of the ansatz. Sec.~\ref{sec:connection_to_mboa} establishes the connection between the present approach and the Moving Born-Oppenheimer approximation (MBOA) of Ref.~\cite{barrera2026MBOA}. Finally,  Secs.~\ref{sec:spinful_particles} and~\ref{sec:bloch_electrons} give supplemental information and numerical parameters for the two model systems discussed in the main text.

\subsection{Self-Consistent, Mixed Quantum-Classical Framework}
\label{sec:self_consistent_framework}

Consider a general Hamiltonian describing a set of slow coordinates $\hatvecx$, with conjugate momenta $\hatvecq$, coupled to fast DOFs denoted generically by a set of operators $\hatvecsigma$,
\begin{equation}
\hat{H}(\hatvecsigma,\hatvecx,\hatvecq)
=
\frac{\hatvecq^2}{2M}
+V(\hatvecx)
+\hat{H}_{\rm int}(\hatvecsigma,\hatvecx).
\label{eq:general_Hamiltonian}
\end{equation}
The total Hilbert space is given by the tensor product of the slow (S) and fast (F) Hilbert spaces, $\mathbb{H} = \mathbb{H}^S\otimes\mathbb{H}^F$.

We carry out so-called \emph{mixed quantum-classical dynamics}~\cite{crespo2018recent}, where the slow DOFs are treated classically while the fast ones remain quantum mechanical. A natural framework to do so is the partial Wigner-Weyl transform~\cite{kapral1999mixed,polkovnikov2010phase,barrera2026MBOA}, which maps quantum-mechanical operators $\hat{\Omega}$ in the full Hilbert space $\mathbb{H}^S\otimes\mathbb{H}^F$ to mixed quantum-classical operators $\weyl{\Omega}(\vecx,\vecq)$ acting only on $\mathbb{H}^F$, with the slow DOFs converted to classical phase space variables $(\vecx,\vecq)$. The quantity $\weyl{\Omega}(\vecx,\vecq)$ is known as the \emph{Weyl symbol} of operator $\hat{\Omega}$, and is defined as,
\begin{equation}
    \weyl{\Omega}(\vecx,\vecq)=\int_{-\infty}^{\infty} \dd \vec{y}\,\bra{\vecx-\vec{y}/2} \hat{\Omega} \ket{\vecx+\vec{y}/2}e^{i\vecq\cdot\vec{y}/\hbar},
    \label{eq:weyl_symbol}
\end{equation}
where $\ket{\vecx}$ denotes the position eigenbasis of $\mathbb{H}^S$. Throughout, bold font denotes such mixed quantum-classical operators. For example, the Weyl symbol of the Hamiltonian Eq.~\eqref{eq:general_Hamiltonian} is,
\begin{equation}
    \weyl{H}(\vecx,\vecq) = (\vecq^2/(2M)+V(\vecx))\,\weyl{I}+\weyl{H}_{\rm int}(\vecx),
    \label{eq:Hamiltonian_weyl_symbol}
\end{equation}
where $\weyl{I}$ denotes the identity operator in $\mathbb{H}^F$. 

It is convenient to combine the position and momentum of the slow DOFs into a single set of phase-space coordinates $\zeta_\alpha=(x_i,q_i)$. Here, Greek indices $\alpha,\beta=1,2,\ldots,2N$ label phase-space coordinates and Latin indices $i,j=1,2,\ldots,N$ label spatial directions. We also introduce the canonical symplectic one-form $\Theta_\alpha=(q_i,0)$, whose phase-space curl gives the canonical symplectic matrix,
$\partial_\alpha\Theta_\beta-\partial_\beta\Theta_\alpha
\equiv J_{\alpha\beta}$, where $J_{x_iq_j} = -\delta_{ij}$ and $J_{x_ix_j}=0$, $J_{q_iq_j}=0$.

\begin{figure}
    \centering
    \includegraphics[]{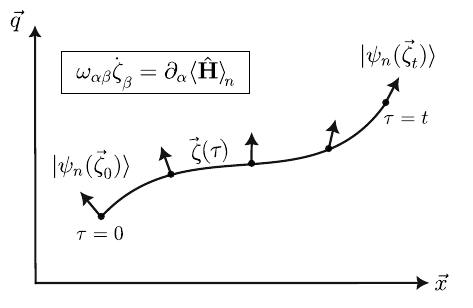}
    \caption{\textit{Self-consistent dynamics of slow classical DOFs coupled to fast quantum DOFs.} As the slow degrees of freedom evolve along some path $\veczeta(\tau)=(\vecx(\tau),\vecq(\tau))$ through their phase space, the fast ones adiabatically follow a variational state $\ket{\smash{\psin(\veczeta(\tau))}}$, for simplicity shown as a spinor, which {need not} be an instantaneous eigenstate of the Hamiltonian $\weyl{H}(\veczeta)$ at the local phase space point. The state $\ket{\smash{\psin(\veczeta)}}$ self-consistently determines the dynamics $\veczeta(\tau)$ through: (i) an energy surface $U^{(n)}(\veczeta)$, (ii) Berry curvatures that enter the symplectic structure of the slow phase space variables. }
    \label{fig:schematic}
\end{figure}

\paragraph{Dynamics of Fast DOFs.---} The key idea is to slave the state of the fast DOFs to the instantaneous configuration of the slow phase space variables $\veczeta=(\vecx,\vecq)$, as shown in Fig.~\ref{fig:schematic}. In particular, we define a manifold of variational states $\ket{\smash{\psiSC_n(\veczeta)}}$ over the slow phase space~\footnote{At this moment, we make no assumptions about the form of the variational states. We return to the question of how to construct them in Sec.~\ref{sec:kinematic_ansatz}.
}. As the slow DOFs evolve along a trajectory $\veczeta(\tau)$, the fast DOFs are assumed to adiabatically follow the state $\ket{\smash{\psiSC_n(\veczeta(\tau))}}$. Note that the variational states need not be the instantaneous eigenstates of the Hamiltonian $\weyl{H}(\veczeta)$ at the local phase space point.

More precisely, consider a trajectory $\veczeta(\tau)$ over the slow phase space. If at $\tau=0$ the fast DOFs are prepared in $\ket{\Psi(0)} = \ket{\smash{\psiSC_n(\veczeta(0))}}$, we assume that the solution to the Schr\"odinger equation, $i\hbar\partial_t\ket{\Psi(t)} = \weyl{H}(\veczeta(t))\ket{\Psi(t)}$, is given in the form,
\begin{equation}
    \ket{\Psi(t)}
    =
    e^{-i \gamma[\veczeta(\tau)]}
    \ket{\smash{\psiSC_n(\veczeta(t))}}.
    \label{eq:slaving_assumption}
\end{equation}
The acquired phase $\gamma[\veczeta(\tau)]$ has both a dynamical and a geometric contribution, and is a functional of the entire trajectory $\veczeta(\tau)$.

The adiabatic gauge potentials (AGPs) of the variational state manifold are given as,
\begin{equation}
    i\hbar\partial_\alpha\ket{\smash{\psiSC_n(\veczeta)}}=\weyl{A}_\alpha\ket{\smash{\psiSC_n(\veczeta)}},
\end{equation}
with $\partial_\alpha \equiv \partial/\partial \zeta_\alpha$. The AGP expectation values
$\mathcal{A}_{\alpha}^{(n)}(\veczeta) =\bra{\psiSC_n}\weyl{A}_\alpha\ket{\psiSC_n} \equiv\langle\weyl{A}_{\alpha}\rangle_n$,
are the Berry connections, and the symmetric and antisymmetric parts of the connected AGP covariance give the quantum metric and Berry curvature tensors,
\begin{equation}
	g_{\alpha\beta}^{(n)}
	=\frac{1}{2}\big\langle\{\AGP{\alpha},\AGP{\beta}\}\big\rangle_{n,c},
	\quad
	\Omega_{\alpha\beta}^{(n)}
	=\frac{i}{\hbar}\big\langle[\AGP{\alpha},\AGP{\beta}]\big\rangle_n.
	\label{eq:geometric_tensors}
\end{equation}
Here, $\langle \weyl{A}\weyl{B}\rangle_{n,c}\equiv\langle \weyl{A}\weyl{B}\rangle_n-\langle \weyl{A}\rangle_n\langle \weyl{B}\rangle_n$ denotes the connected covariance.

In order for the variational ansatz, Eq.~\eqref{eq:slaving_assumption}, to satisfy the Schr\"odinger equation, the slaved states must be instantaneous eigenstates of the {moving Hamiltonian},
\begin{equation}
    E_{\rm mov}^{(n)}(\veczeta,\dot{\veczeta})\ket{\smash{\psiSC_n(\veczeta)}}
    =
    \weyl{H}_{\rm mov}(\veczeta,\dot{\veczeta})\ket{\smash{\psiSC_n(\veczeta)}},
    \label{eq:moving_eigenvalue_eq}
\end{equation}
where, 
\begin{equation}
    \weyl{H}_{\rm mov}(\veczeta,\dot{\veczeta})
    \equiv
    \weyl{H}(\veczeta)
    -
    \dot{\zeta}_\alpha\weyl{A}_\alpha(\veczeta).
    \label{eq:moving_Hamiltonian}
\end{equation}
Equation~\eqref{eq:moving_eigenvalue_eq} follows directly from substituting Eq.~\eqref{eq:slaving_assumption} into the Schr\"odinger equation and using the definition of the AGPs: $i\hbar\partial_t\ket{\smash{\psiSC_n(\veczeta)}}=\dot{\zeta}_\alpha\weyl{A}_\alpha\ket{\smash{\psiSC_n(\veczeta)}}$. Projecting Eq.~\eqref{eq:moving_eigenvalue_eq} onto $\bra{\smash{\psiSC_n(\veczeta)}}$ from the left gives the expression for the moving energy,
\begin{equation}
    E_{\rm mov}^{(n)}(\veczeta,\dot{\veczeta}) = \langle\weyl{H}(\veczeta)\rangle_n - \dot{\zeta}_\alpha\mathcal{A}^{(n)}_\alpha(\veczeta),
\end{equation}
where, $\langle\weyl{H}(\veczeta)\rangle_n \equiv\bra{\smash{\psi_n(\veczeta)}}\weyl{H}\ket{\smash{\psi_n(\veczeta)}}$,
is the average energy, and generates the dynamical contribution to the accumulated phase. In turn, the Berry connection $\mathcal{A}^{(n)}_\alpha(\veczeta)=\langle\psiSC_n(\veczeta)|\weyl{A}_\alpha(\veczeta)|\psiSC_n(\veczeta)\rangle$ generates the geometric contribution to the phase. The total accumulated phase, $\gamma[\veczeta(\tau)]= \frac{1}{\hbar}\int_0^t \d\tau\,E_{\rm mov}^{(n)}\big(\veczeta(\tau),\dot{\veczeta}(\tau)\big)$, is a sum of both of these contributions. Note that $E_{\rm mov}^{(n)}$ plays a role analogous to the quasi-energy in Floquet systems~\cite{schindler2025geometric}.

We emphasize that, up until now, the phase space velocities $\dot{\zeta}_\alpha$ are independent parameters. In the following section, we derive Hamiltonian equations of motion for $\veczeta(\tau)$ which constrain the phase space velocities to be functions of the instantaneous phase space coordinates: $\dot{\zeta}_\alpha=\dot{\zeta}_\alpha(\veczeta)$. Equation~\eqref{eq:moving_eigenvalue_eq} can then be framed as a self-consistent eigenvalue problem at every $\veczeta$.

\paragraph{Dynamics of Slow DOFs.---} Using the time-dependent variational principle~\cite{kramer1981geometry}, the variational states self-consistently determine the dynamics of the slow DOFs through the effective Lagrangian,
\begin{align}
     \mathcal{L}_{\rm eff}^{(n)}(\veczeta,\dot{\veczeta}) &=\Theta_\alpha(\veczeta)\dot{\zeta}_\alpha +\bra{\smash{\psi_n(\veczeta)}}i\hbar\partial_t -\weyl{H}(\veczeta)\ket{\smash{\psi_n(\veczeta)}}, \nonumber \\ 
     &= \Theta_\alpha(\veczeta)\dot{\zeta}_\alpha - \left\langle
    \weyl{H}(\veczeta) -\dot{\zeta}_\alpha\weyl{A}_\alpha(\veczeta) \right\rangle_n.
    \label{eq:lagrangian}
\end{align}
The Euler-Lagrange equations associated to Eq.~\eqref{eq:lagrangian} yield Hamiltonian equations of motion for the slow phase space variables,
\begin{equation}
    \omega^{(n)}_{\alpha\beta}(\veczeta)\,\dot{\zeta}_\beta
    =
    \partial_\alpha\langle\weyl{H}\rangle_n,
    \quad
    {\boldsymbol{\omega}}^{(n)}
    =
    \begin{pmatrix}
        \boldsymbol{\Omega}^{(n)}_{xx}
        &
        -\mathbb{1}+\boldsymbol{\Omega}^{(n)}_{xq}
        \\
        \mathbb{1}+\boldsymbol{\Omega}^{(n)}_{qx}
        &
        \boldsymbol{\Omega}^{(n)}_{qq}
    \end{pmatrix}.
    \label{eq:Hamilton_eom_1}
\end{equation}
Here, $\boldsymbol{\omega}^{(n)}$ is a $2N\times2N$ symplectic matrix. The quantities $\boldsymbol{\Omega}^{(n)}_{ab}$ denote $N\times N$ blocks with matrix elements, $[\boldsymbol{\Omega}^{(n)}_{ab}]_{ij} = \Omega^{(n)}_{a_i b_j}$, where $a,b\in\{x,q\}$, and $\mathbb{1}_{ij} = \delta_{ij}$ denotes the identity matrix. The Berry curvatures of the slaved states therefore modify the symplectic structure of the slow phase space. This is a well-known structure that also arises in the description of wavepacket dynamics in slowly perturbed crystals, where the variational states are chosen to be instantaneous eigenstates of the local Hamiltonian~\cite{sundaram1999wave}. 

\paragraph*{Self-Consistency Condition.---} 
The Hamiltonian equations of motion, Eq.~\eqref{eq:Hamilton_eom_1}, constrain the phase-space velocities to be functions of the local phase-space point, $\dot{\zeta}_\alpha=\dot{\zeta}_\alpha(\veczeta)$. This allows us to combine Eqs.~\eqref{eq:moving_eigenvalue_eq} and~\eqref{eq:Hamilton_eom_1} into a closed set of self-consistent equations for the variational states,
\begin{multline}
        \left( \weyl{H}(\veczeta) - \dot{\zeta}_\alpha(\veczeta)\weyl{A}_\alpha(\veczeta)
        \right)\ket{\smash{\psiSC_n(\veczeta)}} =  E_{\rm mov}^{(n)}(\veczeta)\ket{\smash{\psiSC_n(\veczeta)}},
        \\ 
        \dot{\zeta}_\alpha(\veczeta)
        =
        \big[\omega^{(n)}(\veczeta)\big]^{-1}
        _{\alpha\beta}
        \partial_\beta\langle\weyl{H}(\veczeta)\rangle_n,
    \label{eq:self_consistency}
\end{multline}
Although formally well defined, we do not expect Eq.~\eqref{eq:self_consistency} to be solvable in general. Instead, in this work we use the kinematic states introduced in the main text as our variational ansatz, and take them to be approximate solutions to the self-consistency equations. 

Note that Eq.~\eqref{eq:self_consistency} can be framed as an inverse counterdiabatic problem~\cite{berry2009transitionless,sels2017minimizing}: given the lab frame Hamiltonian $\weyl{H}(\veczeta)$, we seek to construct a moving-frame Hamiltonian $\weyl{H}_{\rm mov}(\veczeta)$ for which $\weyl{H}(\veczeta)$ performs exact transitionless quantum driving. If a solution exists, quantum evolution under the lab frame Hamiltonian is perfectly adiabatic in the basis of eigenstates of $\weyl{H}_{\rm mov}(\veczeta)$.

Combining Eq.~\eqref{eq:Hamilton_eom_1} with the self-consistency condition Eq.~\eqref{eq:self_consistency} yields a more transparent form of the equation of motion. In particular, using the fact that the variational states are eigenstates of $\weyl{H}_{\rm mov} = \weyl{H}-\dot{\zeta}_\beta\weyl{A}_\beta$, application of the Hellmann-Feynman theorem yields,
\begin{align}
    \partial_\alpha\langle\weyl{H}\rangle_n &= \partial_\alpha\left[\langle\weyl{H}-\dot{\zeta}_\beta\weyl{A}_\beta\rangle_n+\dot{\zeta}_\beta\langle\weyl{A}_\beta\rangle_n\right], \nonumber \\
    &= \langle\partial_\alpha\weyl{H}\rangle_n
    +\left(\partial_\alpha\langle\weyl{A}_\beta\rangle_n-\langle\partial_\alpha\weyl{A}_\beta\rangle_n\right)\dot{\zeta}_\beta, \nonumber \\
    &= \langle\partial_\alpha\weyl{H}\rangle_n+\Omega^{(n)}_{\alpha\beta}\dot{\zeta}_\beta.
    \label{eq:Hamiltonian_identity}
\end{align}
Substituting the identity above into Eq.~\eqref{eq:Hamilton_eom_1} gives the relation $\dot{\vecx}=\vecq/M$, and it is possible to recast the dynamics as a second-order differential equation for $\vecx(t)$,
\begin{equation}
\left(\delta_{ij}-{\Omega}_{x_iq_j}^{(n)}\right)\, M\ddot{x}_j
=
-\partial_{x_i}\Un
+{\Omega}_{x_ix_j}^{(n)}\dot{x}_j,
\label{eq:second_order_form}
\end{equation}
Above, the energy surface is defined as $U^{(n)}(\veczeta)=\bra{\smash{\psi_n(\veczeta)}}\weyl{H}_{\rm int}(\vecx)\ket{\smash{\psi_n(\veczeta)}}+V(\vecx)$, and all quantities are evaluated at $\vecq = M\dot{\vecx}$. Note that the momentum Berry curvature $\mathbf{\Omega}_{qq}^{(n)}$ disappears from the second-order form of the dynamics.

\subsection{Variational Ansatz}
\label{sec:kinematic_ansatz}

In this section, we derive the kinematic EOM (Eq.~\eqref{eq:kinematic_eom} of the main text) by using the kinematic states as our variational ansatz. In particular, we postulate,
\begin{equation}
    \ket{\psi_n(\vecx,\vecq)} \approx \ket{\psi_n^{\rm KBO}(\vecx,\vecv = \vecq/M)}. 
    \label{eq:kinematic_ansatz}
\end{equation}
The phase space Berry curvatures are then related to the kinematic Berry curvatures by a simple application of the chain rule,
\begin{equation}
    \boldsymbol{\Omega}^{(n)}_{xx}
    = \boldsymbol{\mathcal{K}}_{xx}^{(n)},  \quad
    \boldsymbol{\Omega}^{(n)}_{xq}
    =
    \frac{\boldsymbol{\mathcal{K}}^{(n)}_{xv}}{M},
    \quad
    \boldsymbol{\Omega}^{(n)}_{qq}
    =
    \frac{\boldsymbol{\mathcal{K}}^{(n)}_{vv}}{M^2}.
\end{equation}
Combining with Eq.~\eqref{eq:second_order_form} gives our main result. Note that using the BO states as a variational ansatz would instead reduce Eq.~\eqref{eq:second_order_form} the standard Eq.~\eqref{eq:adiabatic_eom}. 

A comment is in order regarding the equivalence of the Hamiltonian (Eq.~\eqref{eq:Hamilton_eom_1}) and second-order (Eq.~\eqref{eq:second_order_form}) forms of the equations of motion. The derivation of Eq.~\eqref{eq:second_order_form} assumes that the self-consistency condition (Eq.~\eqref{eq:self_consistency}) holds exactly. When self-consistency is not exactly satisfied, as is the case for our approximate variational ansatz, the Hamiltonian and second-order forms are not equivalent, and will in general produce different dynamics. Here, we do not attempt to establish which form is more accurate. For all examples considered in this work, the two yield equivalent dynamics. The conditions required for this equivalence are discussed in Sec.~\ref{sec:spinful_particles}.

\subsection{Regime of Validity}
\label{sec:superadiabatic_criterion}
In this section, we formulate the superadiabatic criterion determining the regime of validity of Eq.~\eqref{eq:kinematic_eom}. 

The key objects are the superadiabatic AGPs $\weyl{B}_\alpha^{\rm SA}$, defined as the difference between the KBO and BO AGPs,
\begin{equation}
    \weyl{B}_\alpha^{\rm SA} \equiv \weyl{A}_{\alpha}^{\rm KBO}-\weyl{A}_\alpha^{\rm BO}.
\end{equation}
Equivalently, because the BO states are velocity-independent, $\weyl{A}_{x_i}^{\rm KBO}=\weyl{A}_{x_i}^{\rm BO}+\weyl{B}_{x_i}^{\rm SA}$ and $\weyl{A}^{\rm KBO}_{v_i} = \weyl{B}^{\rm SA}_{v_i}$. 

The fast DOFs adiabatically follow kinematic states whenever,
\begin{equation}
    \sum_{m\neq n}\left|\frac{\dot z_\alpha
    \bra{\psi_m^{\rm KBO}}
    \weyl{B}^{\rm SA}_\alpha
    \ket{\psi_n^{\rm KBO}}
    }{
    E^{(n)}_{\rm KBO}-E^{(m)}_{\rm KBO}
    }
    \right|\ll 1.
    \label{eq:kinematic_adiabaticity}
\end{equation}
Above, $E^{(n)}_{\rm KBO}$ are the eigenvalues of $\weyl H^{\rm K}$,
\begin{equation}
    \weyl H^{\rm K}(\vecz)\ket{\psi_n^{\rm KBO}(\vecz)}
    = E^{(n)}_{\rm KBO}(\vecz)\ket{\psi_n^{\rm KBO}(\vecz)},
\end{equation}
not to be confused with the kinematic energy surfaces $\varepsilon_{\rm KBO}^{(n)}$ defined in Eq.~\eqref{eq:KBO_surface}.

We now develop some intuition for when we should expect Eq.~\eqref{eq:kinematic_adiabaticity} to hold. 

Let $\weyl{U}_{\rm BO}(\vecx;\vecx_0)$ denote the unitary transformation defining the BO basis,
\begin{equation}
    \ket{\psi_n^{\rm BO}(\vecx)} = \weyl{U}_{\rm BO}(\vecx;\vecx_0)\ket{\psi_{n}^{\rm BO}(\vecx_0)},
\end{equation}
where $\vecx_0$ denotes an arbitrary reference coordinate. We use primes to denote operators in the frame comoving with the BO basis: $\weyl{O}^\prime \equiv \weyl{U}^\dagger_{\rm BO}\weyl{O}\weyl{U}_{\rm BO}$. For example, by construction, $\weyl{H}_{\rm int}^\prime = \sum_n \varepsilon_{\rm BO}^{(n)}\ket{\psi_n^{\rm BO}}\bra{\psi_n^{\rm BO}}$ is a diagonal operator.

The superadiabatic criterion Eq.~\eqref{eq:kinematic_adiabaticity} can remain controlled even at large velocities, far from BO, provided that $(\weyl{H}^{\rm K})^\prime$ is slowly varying in space. We illustrate this using the concrete example of the ``radial spiral" texture shown in Fig.~\ref{fig:main2}. The BO unitary and its corresponding AGP are chosen as,
\begin{equation}
    \weyl{U}_{\rm BO}(r) = e^{-ikr\weyl{S}_z}, \qquad \weyl{A}_r^{\rm BO} = \hbar k\weyl{S}_z.
\end{equation}
In the co-moving frame, we obtain,
\begin{equation}
    (\weyl{H}^{\rm K})^\prime  = \frac{J}{S}\weyl{S}_x - {\hbar k v_r}\weyl{S}_z.
    \label{eq:H_K_supp}
\end{equation}
In this case, Eq.~\eqref{eq:H_K_supp} becomes perfectly $r$-independent. As such, the fast DOFs adiabatically follow kinematic states even at arbitrary large radial velocities, as shown in Fig.~\ref{fig:main2}, provided the radial acceleration is small. At $v_r=0$, the eigenstates of Eq.~\eqref{eq:H_K_supp} are polarized in the direction of the local magnetic texture, whereas at $v_r\to\infty$ they become polarized in the direction orthogonal to the texture plane. A sufficiently small external force interpolates between these two regimes while satisfying Eq.~\eqref{eq:kinematic_adiabaticity} at all points in time.

\subsection{Connection to the Moving Born-Oppenheimer Approximation}
\label{sec:connection_to_mboa}

A closely related framework, the Moving Born-Oppenheimer approximation (MBOA), was developed in Ref.~\cite{barrera2026MBOA}. The kinematic states follow from the mean-field limit of the MBOA. For clarity, we briefly review the core idea of the MBOA and then illustrate its connection with the present formalism.  

The MBOA begins from the same general Hamiltonian Eq.~\eqref{eq:general_Hamiltonian}, but does not immediately take the Wigner-Weyl transform converting $\hatvecx$ and $\hatvecq$ to phase space variables. Instead, one first performs the BO unitary transformation $\hat{U}_{\rm BO}(\hatvecx)$, constructed to diagonalize the interaction Hamiltonian $\hat{H}_{\rm int}(\hatvecx)$, and only afterward takes the Wigner-Weyl transform. Because these two operations do not commute, the Weyl symbols of operators acquire a different representation in the moving frame (see details in Ref.~\cite{barrera2026MBOA}). In particular, the Weyl symbols of the position and momentum operators transform non-trivially as,
\begin{align}
    \weyl{x}_i&\to\weyl{x}^{\rm M}_i = x_i\,\weyl{I},\label{eq:dressed_position}, \\ 
    \weyl{q}_i&\to\weyl{q}^{\rm M}_i=p_i\,\weyl{I}-\weyl{A}^{\rm BO}_{x_i}(\vecx). \label{eq:dressed_momentum}
\end{align}
Here, the superscript M denotes the Weyl symbols of operators in the moving frame. We refer to these as the ``dressed operators''.

The form of the dressed momentum Eq.~\eqref{eq:dressed_momentum} resembles the minimal-coupling prescription in electromagnetism. The quantity $p_i$ is the canonical momentum, which should be distinguished from the physical momentum operator $\weyl{q}^{\rm M}_i$. Only the expectation value of the latter can be identified with the mass times the mean velocity $\vec{v}$ of the particle. As shown in Ref.~\cite{barrera2026MBOA}, the canonical momentum is instead a \emph{collective} variable of both the fast and slow DOFs.

Whereas the physical momentum contains fluctuations inherited from the fast DOFs, the collective momentum $\vecp$ is almost completely decoupled from this microscopic noise, and behaves as a classical variable that evolves smoothly in time. This physics is encoded in the fact that the dressed momentum $\weyl{q}^{\rm M}$ remains an operator that acts non-trivially on $\mathbb{H}^{F}$. In particular, it can be decomposed into an instantaneous expectation value $\langle \weyl{q}^{\rm M}_i\rangle\equiv Mv_i$ and a covariance tensor $\Sigma_{ij}^q$,
\begin{align}
    Mv_i &\equiv \langle \weyl{q}^{\rm M}_i\rangle = p_i -\langle\weyl{A}_{x_i}^{\rm BO}(\vecx)\rangle, \\
    \Sigma_{ij}^q &\equiv \frac{1}{2}\langle\{\weyl{q}^{\rm M}_i,\weyl{q}^{\rm M}_j\}\rangle_c = \frac{1}{2}\langle\{\weyl{A}^{\rm BO}_{x_i},\weyl{A}^{\rm BO}_{x_j}\}\rangle_c.
    \label{eq:momentum_fluctuations}
\end{align}
Thus, the fluctuations of the physical momentum are directly determined by the variance of the BO AGP. If the average in Eq.~\eqref{eq:momentum_fluctuations} is taken in the instantaneous BO state, $\Sigma_{ij}^q$ reduces to the quantum metric tensor~\cite{barrera2026MBOA}.

The MBOA then assumes that the fast DOFs adiabatically follow the instantaneous eigenstates of the dressed Hamiltonian at fixed $(\vecx,\vecp)$,
\begin{equation}
    \weyl{H}^{\rm M}(\vecx,\vecp)\ket{\psi_n^{\rm MBO}(\vecx,\vecp)} = \mathcal{H}^{\rm MBO}_n(\vecx,\vecp)\ket{\psi_n^{\rm MBO}(\vecx,\vecp)},
    \label{eq:MBO_ansatz}
\end{equation}
where,
\begin{align}
    \weyl{H}^{\rm M}(\vecx,\vecp) &={\left(\vecp-\hat{\vec{\mathbf{A}}}^{\rm BO}_x(\vecx)\right)^2\over 2M}+ V(\vecx)+\weyl{H}_{\rm int}(\vecx)     \label{eq:H_MBOA}
, \\
    &= \weyl{H}(\vecx,\vecp) - \frac{p_i\weyl{A}^{\rm BO}_{x_i}(\vecx)}{M} + \frac{\weyl{A}^{\rm BO}_{x_i}(\vecx)\weyl{A}^{\rm BO}_{x_i}(\vecx)}{2M}\nonumber.
\end{align}
Above, summation over repeated spatial indices $i$ is implied. Note the resemblance of Eqs.~\eqref{eq:MBO_ansatz} and~\eqref{eq:H_MBOA} to the ansatz for the kinematic states Eq.~\eqref{eq:H_K} of the main text. The term linear in the AGP has the same structure, but the MBO Hamiltonian Eq.~\eqref{eq:H_MBOA} contains in addition a term quadratic in the AGP. This quadratic term has two main effects on the dynamics: for the slow DOFs, it reduces in the BO limit to the familiar metric or diagonal Born-Huang correction to the energy surface; for the fast DOFs, it can mediate effective interactions, such as those responsible for the spin squeezing discussed in Ref.~\cite{barrera2026MBOA}. These effects are absent in the methodology developed in Sec.~\ref{sec:self_consistent_framework}.

\paragraph*{Mean-Field Limit.---}
The connection between the two approaches becomes transparent by doing a mean-field approximation to the MBOA. Following Ref.~\cite{barrera2026MBOA}, we decompose the BO AGP into its expectation value and fluctuations $\weyl{A}^{\rm BO}_{x_i}=\langle\weyl{A}^{\rm BO}_{x_i}\rangle+\delta\weyl{A}^{\rm BO}_{x_i}$. Neglecting terms that are quadratic in the fluctuations, the resulting mean-field limit of Eq.~\eqref{eq:H_MBOA} is,
\begin{equation}
    \weyl{H}^{\rm M}_{\rm mf}
    =
    \frac{\vecp^{\,2}}{2M}+V(\vecx)
    -\frac{\langle\hat{{\mathbf{A}}}^{\rm BO}_{x_i}\rangle\langle\hat{{\mathbf{A}}}^{\rm BO}_{x_i}\rangle}{2M}
    +\weyl{H}_{\rm int}(\vecx)
    -v_i\weyl{A}^{\rm BO}_{x_i},
    \label{eq:H_MBOA_mean_field}
\end{equation}
with the mean physical momentum $Mv_i \equiv \langle\weyl{q}^{\rm M}\rangle = p_i-\langle\weyl{A}^{\rm BO}_{x_i}\rangle$. The operator-valued part of Eq.~\eqref{eq:H_MBOA_mean_field} is therefore simply $\weyl{H}_{\rm int}-v_i\weyl{A}^{\rm BO}_{x_i}$. The remaining terms are proportional to the identity on $\mathbb{H}^{F}$ and hence do not affect the instantaneous fast eigenstates. The eigenstates of Eq.~\eqref{eq:H_MBOA_mean_field} therefore coincide with the kinematic states with the identification $q_i=Mv_i$.

This correspondence is justified whenever fluctuations of the BO AGP can be neglected. The mean-field approximation therefore fails when the quantum metric is large or, at finite temperature, when thermal fluctuations of the AGP are non-negligible. In this regime, replacing the physical momentum operator by its mean value $\vecq$ is not justified. Conversely, in systems with many fast DOFs, fluctuations can become negligible in the large-$N$ limit~\cite{barrera2026MBOA}. 

\subsection{Supplementary Information for Example 1: Spinful Particle in a Magnetic Texture}
\label{sec:spinful_particles}

The dynamics shown in Fig.~\ref{fig:main2} are obtained by numerically integrating the mixed TWA Eqs.~\eqref{eq:TWA_fast} and~\eqref{eq:TWA_slow} for a single trajectory with initial condition $\vecx(t=0)=\vecx_0$ and $\dot{\vecx}(t=0)=\vecv_0$. Trajectories are chosen so as to avoid an excluded region of radius $r_{\rm exc} = |\vecv_0|^2/(2kv_\star^2)$, with $v_\star = J/(\hbar Sk)$, where adiabatic following of kinematic states breaks down. 

Fig.~\ref{fig:main2}(a) is obtained for parameters $M=1$, $J=1$, $k=2$, $\hbar S=1$, $V=0$, and initial condition $\vecv_0 = (0,2v_\star)$, $\vecx_0=(5r_{\rm exc},-8r_{\rm exc})$. The spin is initialized in a kinematic state that is maximally polarized anti-parallel to the field, $\ket{\Psi(t=0)} = \ket{\psi_-^{\rm KBO}(\vecx_0,\vecv_0)}$, and the dynamics are integrated up to a final time $t_{\rm final} = 30$.

Fig.~\ref{fig:main2}(c) is obtained for parameters $M=1$, $J=1$, $k=1.5$, $\hbar S=1/2$, with constant external force $V=-0.1x$, and initial condition $\vecx_0=(1,0)$, $\vecv_0=(0,0)$. The dynamics are integrated up to a final time $t_{\rm final}=50$. The two red curves are obtained by initializing the spin maximally polarized parallel $\ket{\psi_+^{\rm KBO}(\vecx_0,\vecv_0)}$, or anti-parallel $\ket{\psi_-^{\rm KBO}(\vecx_0,\vecv_0)}$, to the effective field.

Finally, we note that the analytical expressions for the kinematic Berry curvatures in Eq.~\eqref{eq:radial_spiral_spatial_curvature} follow directly
from the effective field in Eq.~\eqref{eq:radial_spiral_effective_field}
through,
\begin{equation}
    \mathcal{K}_{\alpha\beta}^{(\pm)}
    = \mp \hbar S
    \frac{
    \vec{\mathcal{B}}_{\rm eff}\cdot
    (
    \partial_{z_\alpha}\vec{\mathcal{B}}_{\rm eff}
    \times
    \partial_{z_\beta}\vec{\mathcal{B}}_{\rm eff}
    )
    }{
    |\vec{\mathcal{B}}_{\rm eff}|^3
    }.
\end{equation}

\subsection{Supplementary Information for Example 2: Graphene coupled to an LC Circuit}
\label{sec:bloch_electrons}

The dynamics for the system shown in Fig.~\ref{fig:main3} are characterized by three relevant wavevector scales: (i) the Fermi wavevector $k_F=\varepsilon_F/(\hbar v_D)$, (ii) the displacement from equilibrium $\delta \upkappa =e\Phi_0/(\ell\hbar)$ induced by the circuit flux, and (iii) the characteristic wavevector below which the electric field coupling becomes comparable to the band gap, $\upkappa_\star=\sqrt{\delta\upkappa \omega_{\rm LC}/(2v_D)}$. Here we focus on the regime $\delta\upkappa\sim \upkappa_\star\ll k_F$. Consequently, the electronic distribution is only weakly displaced relative to the Fermi surface, $\delta \upkappa/k_F\ll1$ (see insets in Fig.~\ref{fig:main3}), while non-adiabatic corrections at the Fermi surface remain perturbative, $(\upkappa_\star/k_F)^2\ll1$. In this regime, using Eq.~\eqref{eq:perturbative_eom}, the perturbative limit of the electronic current reads,
\begin{equation}
    \mathcal{J}_{\textrm{KBO},i}^{(n\veck)}
    \approx
    -\partial_{\Phi_i}\eBOnk
    -\kappa^{(n\veck)}_{ij}\ddot{\Phi}_j
    -\Gamma^{(n\veck)}_{ijl}\dot{\Phi}_j\dot{\Phi}_l.
    \label{eq:perturbative_current}
\end{equation}

The clean dynamics considered here require the collisionless regime, $\tau\omega_{\rm LC}\gg1$, where $\tau$ is the relaxation time. In high-mobility graphene, relaxation times of order $\tau\sim1\,\mathrm{ps}$ are experimentally attainable, placing the collisionless regime within reach at THz frequencies. Graphene coupled to LC resonators operating at $\omega_{\rm LC}/2\pi\sim4.5$--$10\,\mathrm{THz}$ has been realized experimentally~\cite{liu2015highly}. For $\varepsilon_F\sim0.1\,\mathrm{eV}$, electric fields of order $E\sim3\,\mathrm{kV/cm}$ are sufficient to obtain $\delta\upkappa \sim \upkappa_\star\sim 0.1k_F$, realizing the hierarchy of momentum scales considered here. 

The dynamics shown in Fig.~\ref{fig:main3} are obtained using a nearest-neighbor tight-binding model of graphene, with hopping amplitude $t=2\hbar v_D/(3a)$, where $a$ is the nearest-neighbor distance. We work in units $\hbar=v_D=a=|e|=\ell=1$, and choose parameters $\varepsilon_F=0.1$, $\omega_{\rm LC}=0.05$, $C=15$, and $L=1/(C\omega_{\rm LC}^2)$. The initial circuit amplitude is fixed by $\upkappa_\star=0.025$, such that $\delta\upkappa=\upkappa_\star=0.25k_F$.

The circuit is initialized with $\vec{\Phi}(0)=(\Phi_0,0)$ and $\dot{\vec{\Phi}}(0)=\vecv_0=(0,v_0)$, where $v_0=2(\upkappa_\star)^2$ and $\Phi_0=v_0/\omega_{\rm LC}$. The electronic occupations are determined from the zero-flux bands at zero temperature, with all states below $\varepsilon_F$ occupied. The pseudospins in
singly occupied momentum sectors are initialized in the lower kinematic eigenstate
$\ket{\psi_-^{\rm KBO}(\veck,\vec{\Phi}(0),\vecv_0)}$. The Brillouin zone is sampled on a $\Gamma$-centered $360\times360$ uniform grid over one primitive reciprocal cell, including both valleys and omitting spin degeneracy. All trajectories are evolved for two bare LC periods, $t_{\rm final}=4\pi/\omega_{\rm LC}$. 

Finally, note that for a constant electric field
$\dot{\vec{\Phi}}=-\ell\vec{E}$, $\ddot{\vec{\Phi}}=0$,
the KBO current in Eq.~\eqref{eq:KBO_current}
can be rewritten in terms of the gauge-invariant wavevector
$\vec{\upkappa}=\veck-e\vec{\Phi}/(\hbar\ell)$ as,
\begin{equation}
    \vec{\mathcal{J}}_{\rm KBO}^{(n\veck)}
    =\frac{e}{\ell}\left[
    \frac{1}{\hbar}\vec{\nabla}_{\upkappa}
    \varepsilon_{\rm KBO}^{(n)}
    -\dot{\vec{\upkappa}}\times
    \vec{\Omega}_{\rm KBO}^{(n)}
    \right],
    \label{eq:wavepacket_velocity}
\end{equation}
where,
\begin{equation}
    \Omega_{{\rm KBO},z}^{(n)}
    =i\left(
    \left\langle
    \partial_{\upkappa_x}\psi_n^{\rm KBO}
    \middle|
    \partial_{\upkappa_y}\psi_n^{\rm KBO}
    \right\rangle
    -\mathrm{c.c.}
    \right),
\end{equation}
is the momentum-space Berry curvature of the kinematic
states, with derivatives taken at fixed $\vecE$.
The expression inside the brackets in Eq.~\eqref{eq:wavepacket_velocity} has the familiar
form of the semiclassical wavepacket velocity, with
the first term the contribution from the band-dispersion 
and the second the anomalous velocity. However,
both the band dispersion and Berry curvature are now
dressed by the electric field. A similar result arises in the perturbative semiclassical theory
of Bloch electrons in uniform electromagnetic fields~\cite{gao2014field}.
Expanding the KBO states to first order in the electric field,
the kinematic Berry curvature is,
\begin{equation}
    \mathcal{K}_{\kappa_i\kappa_j}^{(n)}
    =
    \mathcal{F}_{\kappa_i\kappa_j}^{(n)}
    -e\hbar E_l
    \left(
    \partial_{\kappa_i}G_{jl}^{(n)}
    -\partial_{\kappa_j}G_{il}^{(n)}
    \right)
    +\mathcal{O}(E^2),
\end{equation}
where,
\begin{equation}
    G_{ij}^{(n)}
    =\frac{2}{\hbar^2}\operatorname{Re}
    \sum_{m\neq n}
    \frac{
    \bra{\psi_n^{\rm BO}}A_{\kappa_i}^{\rm BO}\ket{\psi_m^{\rm BO}}
    \bra{\psi_m^{\rm BO}}A_{\kappa_j}^{\rm BO}\ket{\psi_n^{\rm BO}}
    }{\varepsilon_n^{\rm BO}-\varepsilon_m^{\rm BO}}.
\end{equation}
This reproduces the field-induced correction to the Berry curvature
derived in Ref.~\cite{gao2014field}.

\end{document}